\documentclass[useAMS]{mn2e}
\usepackage{times}
\usepackage{graphics}
\usepackage{color}

\begin{document}

\title[A $z=5.78$ Galaxy in Chandra Deep Field-South]{A Star-Forming Galaxy at $z=5.78$ in the Chandra Deep Field-South}
\author[Bunker et al.\ ]{Andrew J.\ Bunker\,$^{1}$\thanks{email:bunker@ast.cam.ac.uk},
Elizabeth R.\ Stanway\,$^{1}$,
Richard S.\ Ellis\,$^{2}$,\\
 \ \ \\
{\LARGE \rm
Richard G.\ McMahon\,$^{1}$,
\& Patrick J.\ McCarthy\,$^{3}$}\\
$^{1}$\,Institute of Astronomy, University of Cambridge,
Madingley Road, Cambridge, CB3\,0HA, U.K.\\ 
$^{2}$\,California Institute of
Technology, Mail Stop 169-327, Pasadena, CA~91109, U.S.A.\\
$^{3}$\,Carnegie Observatories, 813 Santa Barbara St.,
Pasadena, CA~91101, U.S.A.}
\date{
Accepted 2003 March 28.. 
Received 2003 March 25; in original
form 2003 March 09.}

\maketitle

\begin{abstract}
We report the discovery of a luminous $z=5.78$ star-forming galaxy in
the Chandra Deep Field South. This galaxy was selected as an
``$i$-drop'' from the GOODS public survey imaging with {\em HST}/ACS
(object 3 in Stanway, Bunker \& McMahon 2003, astro-ph/0302212). The
large colour of $(i'-z')_{\rm AB}=1.6$ indicated a spectral break
consistent with the Lyman-$\alpha$ forest absorption short-ward of
Lyman-$\alpha$ at $z\approx 6$. The galaxy is very compact (marginally
resolved with ACS with a half-light radius of 0.08\,arcsec, so $r_{\rm
hl}<0.5\,h^{-1}_{70}$\,kpc). We have obtained a deep (5.5-hour)
spectrum of this $z'_{\rm AB}=24.7$ galaxy with the DEIMOS optical
spectrograph on Keck, and here we report the discovery of a single
emission line centred on $8245$\,\AA\ detected at $20\,\sigma$ with a
flux of $f\approx 2\times 10^{-17}\,{\rm ergs\,cm}^{-2}\,{\rm
s}^{-1}$. The line is clearly resolved with detectable structure at
our resolution of better than 55\,km\,s$^{-1}$, and the only plausible
interpretation consistent with the ACS photometry is that we are
seeing Lyman-$\alpha$ emission from a $z=5.78$ galaxy. This is the
highest redshift galaxy to be discovered and studied using {\em HST}
data. The velocity width ($\Delta v_{\rm FWHM}=260\,{\rm km\,s}^{-1}$)
and rest-frame equivalent width ($W_{\rm rest}^{\rm Ly\alpha}=20$\,\AA
) indicate that this line is most probably powered by star formation,
as an AGN would typically have larger values.  The starburst
interpretation is supported by our non-detection of the
high-ionization N{\scriptsize~V}\,$\lambda$\,1240\,\AA\ emission line,
and the absence of this source from the deep {\em Chandra} X-ray
images. The star formation rate inferred from the rest-frame UV
continuum is $34\,h_{70}^{-2}\,M_{\odot}\,{\rm yr}^{-1}$
($\Omega_M=0.3$, $\Omega_{\Lambda}=0.7$). This is the most luminous
starburst known at $z>5$. Our spectroscopic redshift for this object
confirms the validity of the $i'$-drop technique of Stanway, Bunker \&
McMahon (2003) to select star-forming galaxies at $z\approx 6$.

\end{abstract}
\begin{keywords}
galaxies: evolution --
galaxies: formation --
galaxies: starburst --
galaxies: individual: SBM03\#3 --
galaxies: high redshift --
ultraviolet: galaxies
\end{keywords}

\section{Introduction}
\label{sec:intro}

A key question in modern cosmology is the star formation rate of
galaxies in the young Universe. In the past few years, the Lyman break
technique (Steidel, Pettini \& Hamilton 1995) has proved very
successful in identifying star-forming galaxies at $z=3-4$, through
their rest-frame UV continuum absorbed by the Lyman-$\alpha$ forest of
the intergalactic medium at $\lambda_{\rm rest}<1216$\,\AA\ (e.g.,
Steidel et al.\ 1996; Steidel et al.\ 1999). In a recent paper
(Stanway, Bunker \& McMahon 2003) we have pushed this technique to
even earlier epochs, using the latest public data from the `Great
Observatories Origins Deep Survey' (GOODS; Dickinson \& Giavalisco
2002), taken with the new Advanced Camera for Surveys (ACS; Ford et
al.\ 2002) on the {\em Hubble Space Telescope (HST)}. By selecting on
extreme colours between the F775W $i'$ and F850LP $z'$ filters, we
identify galaxies in the Chandra Deep Field-South (CDF-S) with
$(i'-z')_{\rm AB}>1.5$ which are likely to be at redshifts
$5.6<z<6.5$. In this {\em Letter} we report on our spectroscopic
confirmation with Keck/DEIMOS of one of our candidates -- object 3 in
Stanway, Bunker \& McMahon (2003), hereafter called SBM03\#3. We show
that this $z'_{\rm AB}=24.7$ galaxy has Lyman-$\alpha$ emission at a
redshift of $z=5.78$, and does indeed lie within the expected redshift
range. This confirms the validity of $i'$-drop technique of Stanway,
Bunker \& McMahon (2003).

\section{Observations and Data Reduction}
\label{sec:obs}

We observed SBM03\#3 ($\alpha_{2000}=03^{h}32^{m}\,25.59^{s}$,
$\delta_{2000}=-27^{\circ}\,55^{\prime}\,48.4^{\prime\prime}$;
Stanway, Bunker \& McMahon 2003) on the nights of U.T. 2003 January
8\,\&\,9, using the new Deep Imaging Multi-Object Spectrograph
(DEIMOS; Davis et al.\ 2002, Phillips et al.\ 2002) at the Cassegrain
focus of the 10-m Keck{\scriptsize~II} Telescope.  DEIMOS has 8 MIT/LL
$2k\times 4k$ CCDs with $15\,\mu$m pixels and an angular pixel scale
of 0.1185\,arcsec\,pix$^{-1}$.  We used a slitmask to target various
objects in the CDF-S over the $16.5\times5$\,arcmin DEIMOS field, as
part of an ongoing program to obtain spectra of objects with extreme
optical/near-infrared colours (to be reported elsewhere). Each slitlet
was 1\,arcsec wide, and the slit containing SBM03\#3 was at the
extreme edge of the field, so although the slit was 100\,arcsec long,
the target was only 3\,arcsec from the slit edge. The observations
were obtained using the Gold 1200\,line\,mm$^{-1}$ grating in first
order blazed at $\lambda_{\rm blaze} = 7500$\,\AA , producing a
dispersion of $0.320$\,\AA\,pixel$^{-1}$.  The grating was tilted to
sample the wavelength range $\lambda\lambda_{\rm obs}\,7100-9700$\,\AA
, corresponding to a rest-frame wavelength of $\lambda\lambda_{\rm
rest}\,1047-1430$\,\AA\ at the redshift of SBM03\#3 ($z=5.78$, see
Section~\ref{subsec:redshift}). A small region in the middle of the
wavelength range $\lambda\lambda_{\rm obs}\,8405-8412$\,\AA\ is
unobserved as it falls in the gap between two CCDs. We used the OG550
order-blocking filter to remove all light at wavelengths short-ward of
5500\,\AA , so we should not be affected by second-order light.

The spectral resolution was measured to be $\Delta\lambda_{\rm
FWHM}^{\rm obs}\approx 1.35-1.50$\,\AA\ from the sky lines and those of
reference arc lamps ($\Delta v_{\rm FWHM}= 55\,{\rm km\,s}^{-1}$, a
resolving power of $\lambda\,/\,\Delta\lambda_{\rm FWHM}=5500$). As the
seeing disk (typically 0.7\,arcsec FWHM) was smaller than the slit width
of 1.0\,arcsec, the true resolution is somewhat better for a source
which does not fill the slit.

The observations spanned an airmass range of $1.4-1.6$. The PA of slits
on the mask was 6\,deg, chosen to be close North--South (0\,deg, the
parallactic angle) as the field was observed during meridian transit.  A
total of 20\,ksec of on-source integration was obtained, and this was
broken into individual exposures each of duration 2400\,s to enable more
effective cosmic ray rejection. We dithered the telescope 1.5\,arcsec
along the slit between integrations. The spectrophotometric standard
star HZ\,44 (Massey et al.\ 1988; Massey \& Gronwall 1990) was observed
to determine the sensitivity function. The flux calibration was checked
with the spectra of the five alignment stars of known broad-band
photometry ($I\approx 17-19$\,mag), used to position the CDF-S mask
through $2''\times2''$ alignment boxes.

The data reduction followed standard procedures using {\em IRAF}.
Each frame first had the bias subtracted, determined from the overscan
region appropriate to that CCD. Contemporaneous flat fields were
obtained with a halogen lamp immediately after the science exposures,
and these internal flats were normalized through division by the
extracted lamp spectrum, and corrected for non-uniform illumination of
the slit by comparison to twilight sky and dome spectral flat
fields. Wavelength calibration was obtained from Ne$+$Ar$+$Hg$+$Kr
reference arc lamps, and a cubic fit to the centroids of $\approx 20$
arc lines for each CCD created a wavelength solution with {\em rms}
residuals of 0.02\,\AA. The wavelength calibration was checked with
the centroids of prominent sky lines. The spectrum was then rectified
(straightening the sky lines), and sky subtraction was performed by a
fifth-order polynomial fit to each rectified detector column (parallel
to the slit), excluding from the fit those regions occupied by
sources.

\section{Results}
\label{sec:results}

\subsection{The redshift of SMB03\#3}
\label{subsec:redshift}

\begin{figure}
\resizebox{0.49\textwidth}{!}{\includegraphics{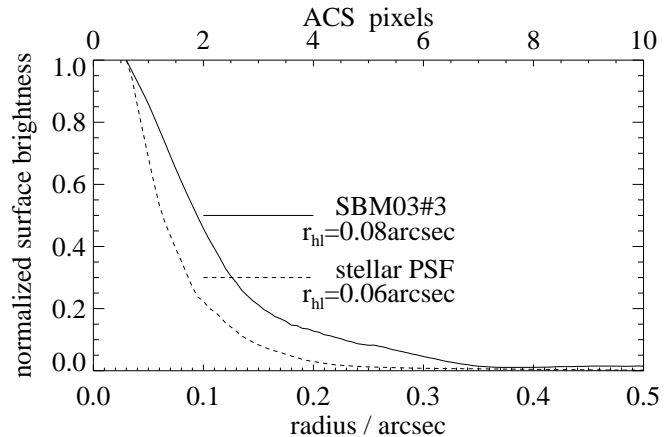}}
\caption{The normalized radial surface brightness profile of SBM03\#3
from the {\em HST}/ACS F850LP $z'$-band image, compared to that of a
star. The high-$z$ galaxy is only marginally resolved.}
\label{fig:radprof}
\end{figure}

The combined 5.5-hours of DEIMOS spectroscopy showed a clear emission
line detected at $20\,\sigma$ at the location in the slit
corresponding to the expected position of galaxy SBM03\#3. The peak of
the line emission is at $\lambda_{\rm obs}= 8245.4\pm 0.9$\,\AA , and
the integrated flux is within 30 per cent of $f=2\times 10^{-17}\,{\rm
ergs\,cm}^{-1}\,{\rm s}^{-1}$, extracting over 17 pixels (2\,arcsec)
and measuring between the zero-power points ($8242-8260$\,\AA ). This
is actually a lower limit on the line flux because of possible slit
losses if the emission line is spatially extended, although this is
probably negligible since the galaxy is very compact
(Fig.~\ref{fig:radprof}).  The flux measurement is consistent over
both nights.  The emission line is not seriously contaminated by
residuals from strong OH lines but does straddle a weak complex of sky
lines. Fig.~\ref{fig:Lya2D} shows the combined 2-dimensional long-slit
spectrum around this emission line.

Single-line redshifts are often open to question, but the most
probable identification of this solo emission line is
Lyman-$\alpha$\,$\lambda$\,1215.8\,\AA\ at $z=5.7825\pm 0.001$, given
that this galaxy was pre-selected as an $i'$-drop to have a
photometric redshift of $z\approx 6$. We briefly consider (and rule
out) other possible redshift assignations. This line emission is
extremely unlikely to be H$\beta$\,$\lambda$\,4861.3\,\AA\ or
[O{\scriptsize~III}]\,$\lambda\lambda$\,5006.8,4958.9\,\AA\ at
$z\approx 0.7$ as we do not detect the other nearby lines in this
complex. If this line emission were
[O{\scriptsize~II}]\,$\lambda\lambda$\,3726.1,3728.9\,\AA\ at
$z=1.212$, we would expect to resolve comfortably the
[O{\scriptsize~II}] doublet with our spectral resolution of $55\,{\rm
km\,s}^{-1}$, as demonstrated by Davis et al.\ (2002) using DEIMOS in
the same configuration.  The most probable interpretation is that we
are seeing Lyman-$\alpha$\,$\lambda$\,1215.8\,\AA\ at $z=5.78$. The
only plausible alternative to high-redshift Lyman-$\alpha$ is
H$\alpha$\,$\lambda$\,6562.8\,\AA\ at $z=0.256$: however, the
H$\alpha$ interpretation would not produce the large flux decrement
observed between the $z'$-band and $i'$-band. Hence we conclude that
Lyman-$\alpha$ emission at $z=5.78$, associated with the $i'$-drop
galaxy SBM03\#3, is by far the most probable identification of the
line (see Stern et al.\ 2000 for a discussion of one-line redshift
determinations for high-$z$ galaxies).

\subsection{Spectral Profile of the Lyman-$\alpha$ Emission}
\label{subsec:lineprof}

\begin{figure*}
\resizebox{0.66\textwidth}{!}{\includegraphics{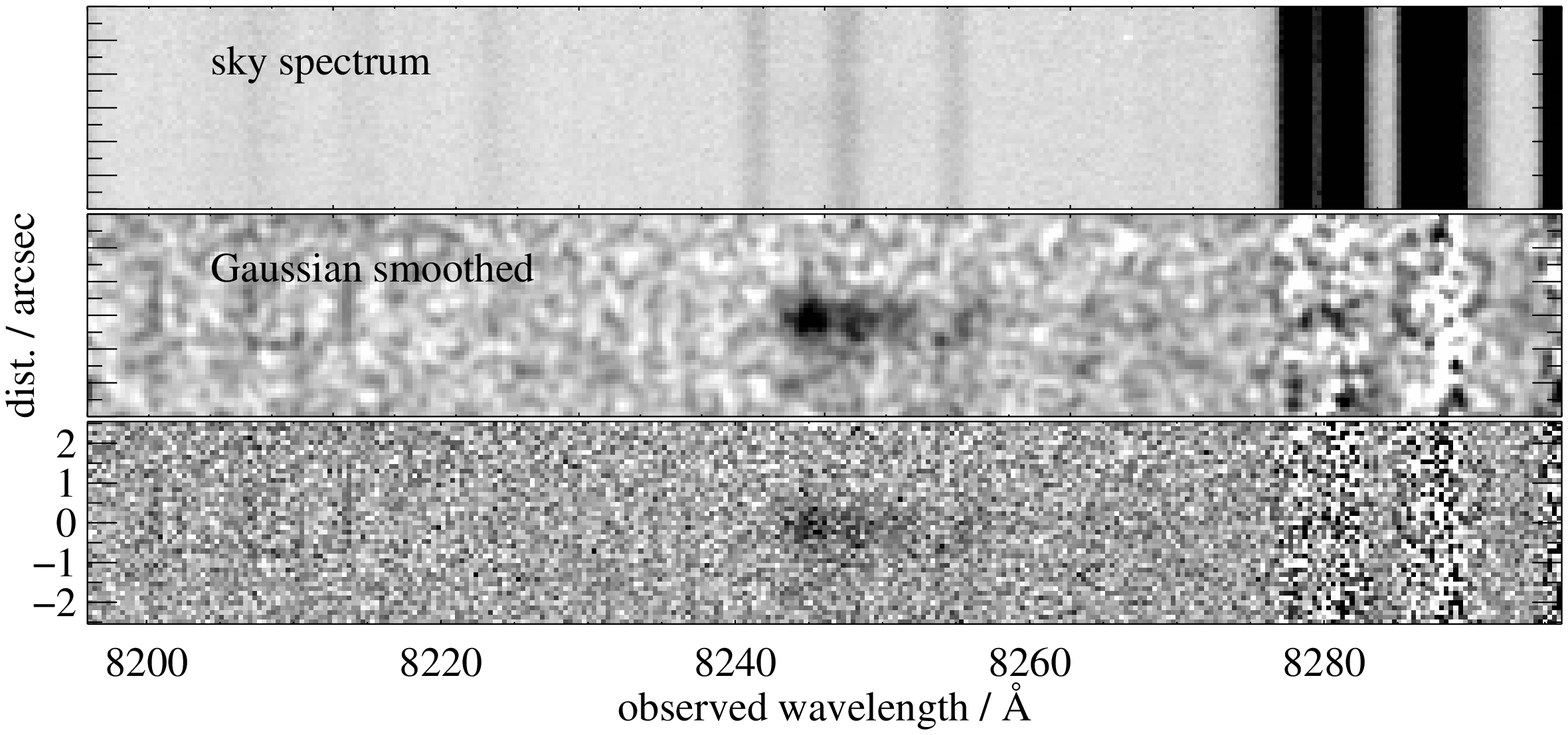}}
\resizebox{0.33\textwidth}{!}{\includegraphics{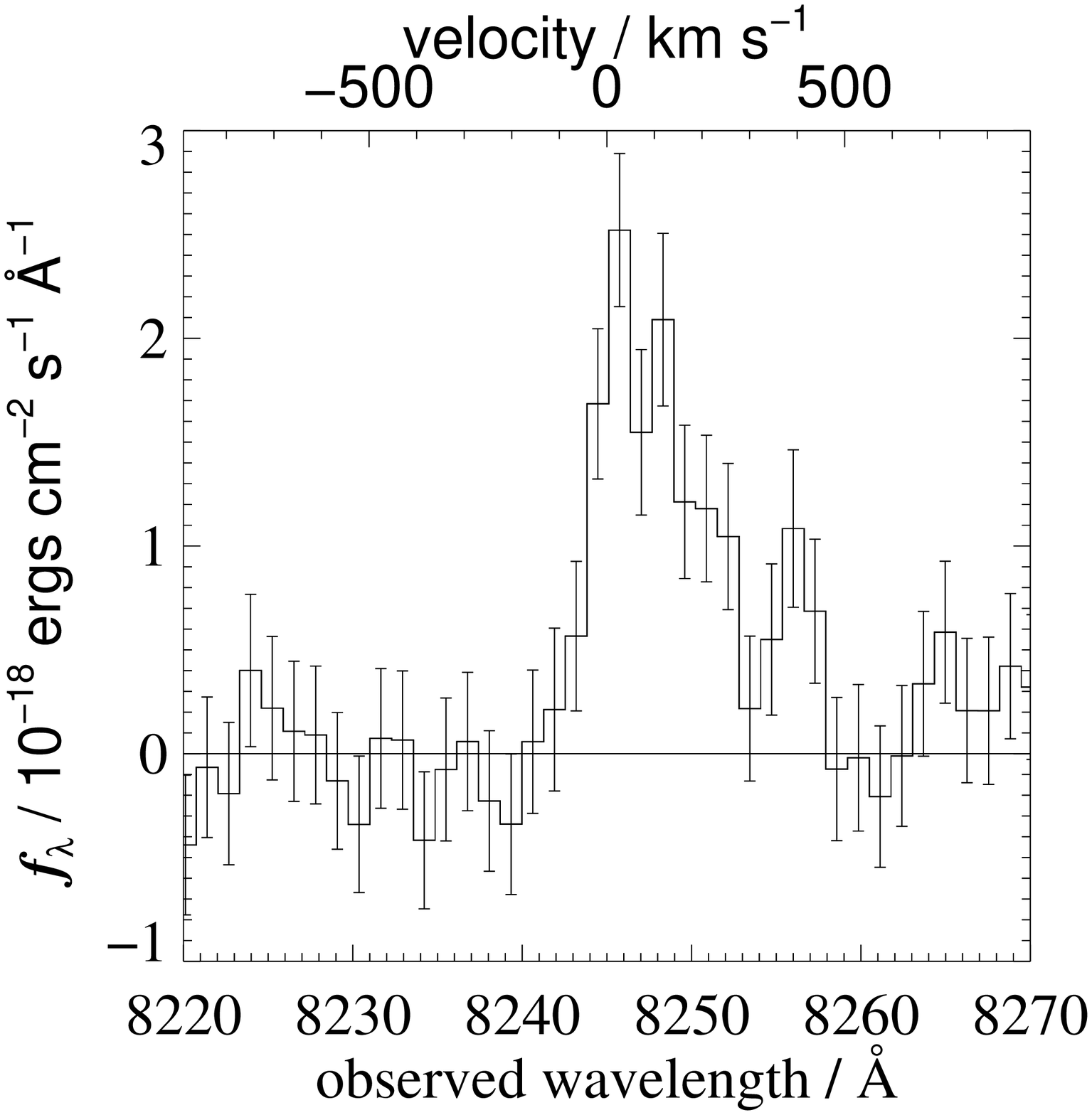}}
\caption{(a) The lower panel shows the 2D spectrum around
Lyman-$\alpha$ (100\AA\ by 5arcsec), the middle panel shows this
smoothed with a Gaussian of $\sigma=1$\,pix, and the upper panel shows
the sky spectrum for this wavelength range.(b) The 1D spectrum around
Lyman-$\alpha$, extracted over a 9\,pixel (1\,arcsec) width. The data
have been binned into independent resolution elements of 1.3\,\AA\
(4\,pixels).}
\label{fig:Lya2D}
\label{fig:Lya1D}
\end{figure*}

The Lyman-$\alpha$ line is clearly resolved spectrally at this
$55\,{\rm km\,s}^{-1}$ resolution, with significant velocity structure
evident (Fig.~\ref{fig:Lya1D}). There appears to be a weak second peak
of emission slightly to the red of the main Lyman-$\alpha$ emission,
around 8256\,\AA , and there is more marginal evidence for narrow
absorption at the core of the main emission line.

The spectral width of the Lyman-$\alpha$ line is $(7.3\pm
0.5)$\,pixels FWHM, excluding second peak at 8256\,\AA . This is
equivalent to a velocity width of $\Delta v_{\rm FWHM}=(260\pm
20)\,{\rm km\,s}^{-1}$ after deconvolution with the instrumental
width.  The line width is similar to those of other $z\approx 6$
galaxies (e.g., Rhoads et al.\ 2003; Kodaira et al.\ 2003; Lehnert \&
Bremer 2003). The Lyman-$\alpha$ profile is asymmetric, with a
pronounced red wing but a sharper decline in flux density on the blue
side (Fig.~\ref{fig:Lya1D}). This appears to be a common feature in
high-$z$ starbursts with Lyman-$\alpha$ in emission (e.g., Dey et al.\
1998; Lowenthal et al.\ 1997; Bunker, Moustakas \& Davis 2000; Ellis
et al.\ 2001), and is most likely due to an outflow of neutral
hydrogen, where we only see the back-scattered Lyman-$\alpha$ from the
far side of the expanding nebula -- only the photons on the red side
of the resonant Lyman-$\alpha$ emission-line profile can escape, with
the blue wing being absorbed by neutral gas (within the galaxy and in
the Lyman-$\alpha$ forest). The P\,Cygni-like profile of
Lyman-$\alpha$ is consistent with this outflow model. Hence the
profile of Lyman-$\alpha$ is skewed to the red, so the systemic
redshift determined from the peak of the emission may be
overestimated.  Near-infrared spectroscopy by Pettini et al.\ (2001)
of the rest-frame optical forbidden lines and Balmer lines in
$z\approx 3$ galaxies shows a $200-1100\,{\rm km\,s}^{-1}$ redshifting
of the Lyman-$\alpha$ relative to the nebular emission lines.

The velocity dispersion of $\sigma_{\rm LOS} = (110\pm 10)\,{\rm
km\,s}^{-1}$ inferred from the line-width of Lyman-$\alpha$ is likely
to be a lower limit, as the line profile is truncated. However, the
velocity width of Lyman-$\alpha$ cannot be reliably used to estimate
the galaxy mass as it is unlikely to be representative of the true
velocity dispersion, since resonant scattering will broaden the
spectral profile of the escaping photons (e.g., Binette et al.\ 1993).

\subsection{AGN or Starburst?}
\label{subsec:AGNorStarburst}

Careful inspection of the spectrum did not reveal any lines at other
wavelengths in the same spatial location as the Lyman-$\alpha$
emission. The only other significant line which should fall within our
spectral coverage (although close to the gap between the CCDs) is the
high-ionization rest-UV doublet
N{\scriptsize~V}\,$\lambda\lambda$\,1238.8,1242.8\,\AA, which is
usually prominent in active galactic nuclei (AGN). Our flux limits at
8402\,\&\,8429\,\AA\ are $f<2\times 10^{-18}\,{\rm
ergs\,cm}^{-2}\,{\rm s}^{-1}$ ($3\,\sigma$), conservatively assuming
the lines are extended and that half the light is lost in gap between
the CCDs at $8405-8412$\,\AA . Our lower limit of $f({\rm
Ly\alpha})/f({\rm N{\scriptsize~V}})>10$ ($3\,\sigma$) compares with
the typical line ratios from composite QSO spectra of $f({\rm
Ly\alpha})/f({\rm N{\scriptsize~V}})=4.0$ (Osterbrock 1989). Hence,
the non-detection of N{\scriptsize~V}\,1240\,\AA\ favours the
interpretation that the Lyman-$\alpha$ arises from the Lyman continuum
flux produced by OB stars, rather than the harder UV spectrum of a
QSO.  The velocity width of Lyman-$\alpha$ ($260\,{\rm km\,s}^{-1}$
FWHM) is also significantly narrower than seen in the broad-line
regions of AGN, even after correcting for the self-absorbed blue wing
(\S\,\ref{subsec:lineprof}), again suggesting a starburst rather than
an AGN.  SBM03\#3 is undetected in the Chandra X-ray map of the CDF-S
(Giacconi et al.\ 2002) to $2\,\sigma$ limits of $5.5\times
10^{-17}\,{\rm ergs\,cm}^{-2}\,{\rm s}^{-1}$ and $4.5\times
10^{-16}\,{\rm ergs\,cm}^{-2}\,{\rm s}^{-1}$ in the soft (0.5--2\,keV)
and hard (2--10\,keV) X-ray bands respectively. We note that several
other Lyman-$\alpha$ emitters at high redshift are also undetected in
X-rays (Malhota et al.\ 2003).

\subsection{Continuum Shape and Lyman-$\alpha$ Equivalent Width}
\label{subsec:EquivWidth}

We do not have a significant detection of the the continuum in the
Keck/DEIMOS spectroscopy, but the flux measurement of $z'_{\rm
AB}=24.67\pm 0.03$ from the {\em HST}/ACS F850LP $z'$-band image
--which encompasses the redshifted Lyman-$\alpha$ line at its blue
end-- enables us to determine the equivalent width of this line and
the continuum flux density ($L_{\nu}=2.7\,h^{-2}_{70}\times
10^{29}\,{\rm ergs\,s}^{-1}\,{\rm Hz}^{-1}$ at $\lambda_{\rm
rest}\approx 1300$\,\AA , where we assume $\Omega_{\rm M}=0.3$ and
$\Omega_{\Lambda}=0.7$ throughout). The emission line has a luminosity
$L=(7\pm 2)\times 10^{42}\,h^{-2}_{70}\,{\rm ergs\,s}^{-1}$ and
accounts for only $\approx 4$ per cent of the $z'$-band flux and $\sim
25$ per cent of the $i'$-band flux, and once the effect of this line
contamination of the broad-band magnitudes is removed, the equivalent
width of the line is $W_{\rm obs}^{\rm Ly\alpha}=(140\pm 50)$\,\AA\ in
the observed frame, assuming that there is negligible continuum flux
below 8245\,\AA\ (short-ward of $\lambda_{\rm rest}1216$\,\AA , the
Lyman-$\alpha$ line) due to absorption by the Lyman-$\alpha$ forest.

The rest-frame equivalent width at $z=5.78$ is $W_{\rm rest}^{\rm
Ly\alpha}=(20\pm 7)$\,\AA\ which is within the realm of what is
observed in in star forming galaxies. From stellar synthesis models of
star-forming regions (e.g., Charlot \& Fall 1993), the theoretical
Lyman-$\alpha$ equivalent width for a young region of active star
formation is $W_{\rm rest}^{\rm Ly\alpha}\approx 100-200$\,\AA
. However, the observed Lyman-$\alpha$ emission from star-forming
galaxies is invariably much weaker, typically $W_{\rm
rest}=5-30$\,\AA\ (e.g., Steidel et al.\ 1996; Warren \& M\o ller
1996), or even in absorption. The rest-frame equivalent width for
SBM03\#3 lies at the upper end of this observed range and is similar
to the narrow-band selected galaxies at $z>5.5$ (Hu, McMahon \& Cowie
1999; Hu et al.\ 2002; Kodaira et al.\ 2003).

From the $z'$-band, the UV continuum implies an unobscured star
formation rate of $33.8\,h^{-2}_{70}\,M_{\odot}\,{\rm yr}^{-1}$
(Stanway, Bunker \& McMahon 2003). The star formation rate inferred
from the Lyman-$\alpha$ emission would be $\approx 5$ times less than
this, probably due to selective extinction of this line through
resonant scattering. SBM03\#3 has the highest UV luminosity (i.e., the
largest unobscured star formation rate) of any starburst yet found at
$z>5.5$.

We obtained 10\,ksec of near-infrared imaging with a $K_{\rm s}$
filter ($\lambda_{\rm cent}=2.15\,\mu$, equivalent to $\lambda_{\rm
rest}=3200$\,\AA ) using the WIRC camera on the Las Campanas 2.5-m du
Pont telescope on the night of 2003 February 20 U.T. The seeing was
0.7\,arcsec, oversampled by the 0.1\,arcsec pixels of the $1024^2$
Rockwell HgCdTe array. The galaxy SBM03\#3 was undetected, with a
$2\,\sigma$ limiting magnitude of $K_{\rm s}>20.6$ (Vega
magnitudes). If the spectrum long-ward of Lyman-$\alpha$ is a power
law, we can contrain the slope to be $\alpha<0.4$ (where
$f_{\lambda}\propto \lambda^{\alpha}$) from $(z'_{\rm AB}-K_{\rm
s})<4.1$ ($2\,\sigma$).

Knowing the redshift and the contamination of the broad-band
magnitudes by the line emission, the continuum depression as inferred
from the $(i'-z')_{\rm AB}=1.60\pm 0.13$ colour may be used to make a
direct estimate of $D_A$, the absorption due to intervening
cosmological H{\scriptsize~I} clouds at $z\approx 4.8-5.8$. Outflowing
neutral hydrogen intrinsic to the source may also play a significant
role in absorbing the blue-wing of Lyman-$\alpha$.  Formally, the
$D_{A}$ continuum break (Oke \& Korycansky 1982) is defined as
\begin{equation}
D_{A}=\left( 1 - \frac{f_{\nu}(1050-1170\,{\rm \AA})_{\rm
obs}}{f_{\nu}(1050-1170\,{\rm \AA})_{\rm pred}}\right) 
\end{equation}
(e.g., Schneider, Schmidt \& Gunn 1991; Madau 1995). The {\em HST}/ACS
F775W $i'$-band samples the rest-frame at $1030-1220$\,\AA , and the
measured magnitude is $i'_{\rm AB}=26.27\pm 0.13$ (Stanway, Bunker \&
McMahon 2003). After correcting for the fraction of the $i'$- and
$z'$-filters which lie above the redshifted 1216\,\AA\ break, we
calculate $D_{A}=0.95^{+0.05}_{-0.10}$ assuming an intrinsic spectral
shape of $f_{\lambda}\propto \lambda^{-1.1}$, the average for the $z\approx
3$ Lyman break galaxies (Meurer et al.\ 1997). This is consistent with
the values derived from SDSS quasars at similar redshifts by Fan et
al.\ (2001).

\subsection{The size of SBM03\#3}
\label{subsec:Whatis}

The galaxy SBM03\#3 is very compact in the {\em HST} images
(Fig.~\ref{fig:radprof}) -- in the highest S/N ACS $z'$-image it is
only marginally resolved with a half-light radius of
$r_{\rm hl}=0.08$\,arcsec (Stanway, Bunker \& McMahon 2003), compared to
$r_{\rm hl}=0.06$\,arcsec for unsaturated stars in the ACS image. Hence,
the half-light radius of the star-forming region at $z=5.78$ must be
$r_{\rm hl}\ll 0.5\,h^{-1}_{70}$\,kpc. This is very compact, and
corresponds the the physical size of dwarf spheroidals at lower
redshift. SBM03\#3 is more compact than typical Lyman-break galaxies
at $z\approx 3$: Lowenthal et al.\ (1997) find a median
$r_{\rm hl}=3.5\,h_{70}^{-1}$\,kpc and a range $1.7 < r_{\rm hl} <
7\,h^{-1}_{70}$\,kpc. We have also compared the spatial extent of the
line emission with a cross-cut of an alignment star in the same
slitmask (i.e., identical seeing and airmass): the Lyman-$\alpha$ line
has a very compact spatial extent, unresolved in $\approx 0.7$\,arcsec
FWHM seeing, so this seems not to be one of the cases where the
Lyman-$\alpha$ emission line morphology is significantly more extended
than the continuum (e.g., Steidel et al.\ 2000; Bunker, Moustakas \&
Davis 2000).

\section{Conclusions}

We have obtained deep spectroscopy with Keck/DEIMOS of SBM03\#3, an
$i'$-drop in the Chandra Deep Field-South photometrically-selected
from {\em HST}/ACS images to lie at $z\approx 6$. We discover a single
emission line with peak intensity at 8245\,\AA , consistent with
Lyman-$\alpha$ emission form a galaxy at $z=5.78$.  The
spectrally-resolved profile of the emission line is asymmetric (as
high-$z$ Lyman-$\alpha$ tends to be) with a P\,Cygni-like profile and
a sharp cut-off on the blue wing.  The line flux is $\approx 2\times
10^{-17}\,{\rm ergs\,cm}^{-2}\,{\rm s}^{-1}$. The equivalent width
inferred from the $z'$-band photometry from {\em HST}/ACS is $W_{\rm
rest}=20$\,\AA\ , which is within the range seen in high-$z$
star-forming galaxies. The galaxy is undetected in X-rays by Chandra,
and the velocity width of the Lyman-$\alpha$ is comparatively narrow
($v_{\rm FWHM}=260\,{\rm km\,s}^{-1}$) and we do not detect the
high-ionization line N{\scriptsize~V}\,$\lambda$\,1240\,\AA , all of
which support the view that this line emission is powered by star
formation rather than an AGN. Our spectroscopic redshift for this
object confirms the validity of the $i'$-drop selection technique of
Stanway, Bunker \& McMahon (2003) to select star-forming galaxies at
$z\approx 6$.

The relatively bright rest-frame UV continuum flux and Lyman-$\alpha$
line luminosity of SBM03\#3 and the recent WMAP results which indicate
the epoch of reionization to lie at $z>10$ (Kogut et al.\ 2003) bodes
well for searches for UV luminous star-forming galaxies at $z>7$ such as
those which will be possible using the narrow-band near-IR imaging
system DAZLE (the Dark Ages `Z' Lyman Explorer; McMahon et al.\ in
preparation), {\em HST}/WFC3 and NGST.

\subsection*{Acknowledgements}

We are extremely grateful for the help and support we received while
observing at Keck, and in particular thank Greg Wirth, Bob Goodrich and
Chuck Sorenson. We used Drew Phillips' extremely useful ``dsimulator''
software for slitmask design.
We have had useful discussions on the
reduction of optical slitmask spectroscopy with Daniel Stern, Adam
Stanford and Alison Coil.  Some of the data presented herein were
obtained at the W. M.\ Keck Observatory, which is operated as a
scientific partnership among the California Institute of Technology, the
University of California and the National Aeronautics and Space
Administration.  The Observatory was made possible by the generous
financial support of the W. M.\ Keck Foundation.
This paper is based on
observations made with the NASA/ESA Hubble Space Telescope, obtained
from the Data Archive at the Space Telescope Science Institute, which is
operated by the Association of Universities for Research in Astronomy,
Inc., under NASA contract NAS 5-26555. 
We
are grateful to the GOODS team for making their reduced images public.
ERS acknowledges a Particle Physics and
Astronomy Research Council (PPARC) studentship supporting this study.

\bsp

\begin{figure*}
\begin{minipage}{0.83\textwidth}
\resizebox{\textwidth}{!}{\includegraphics{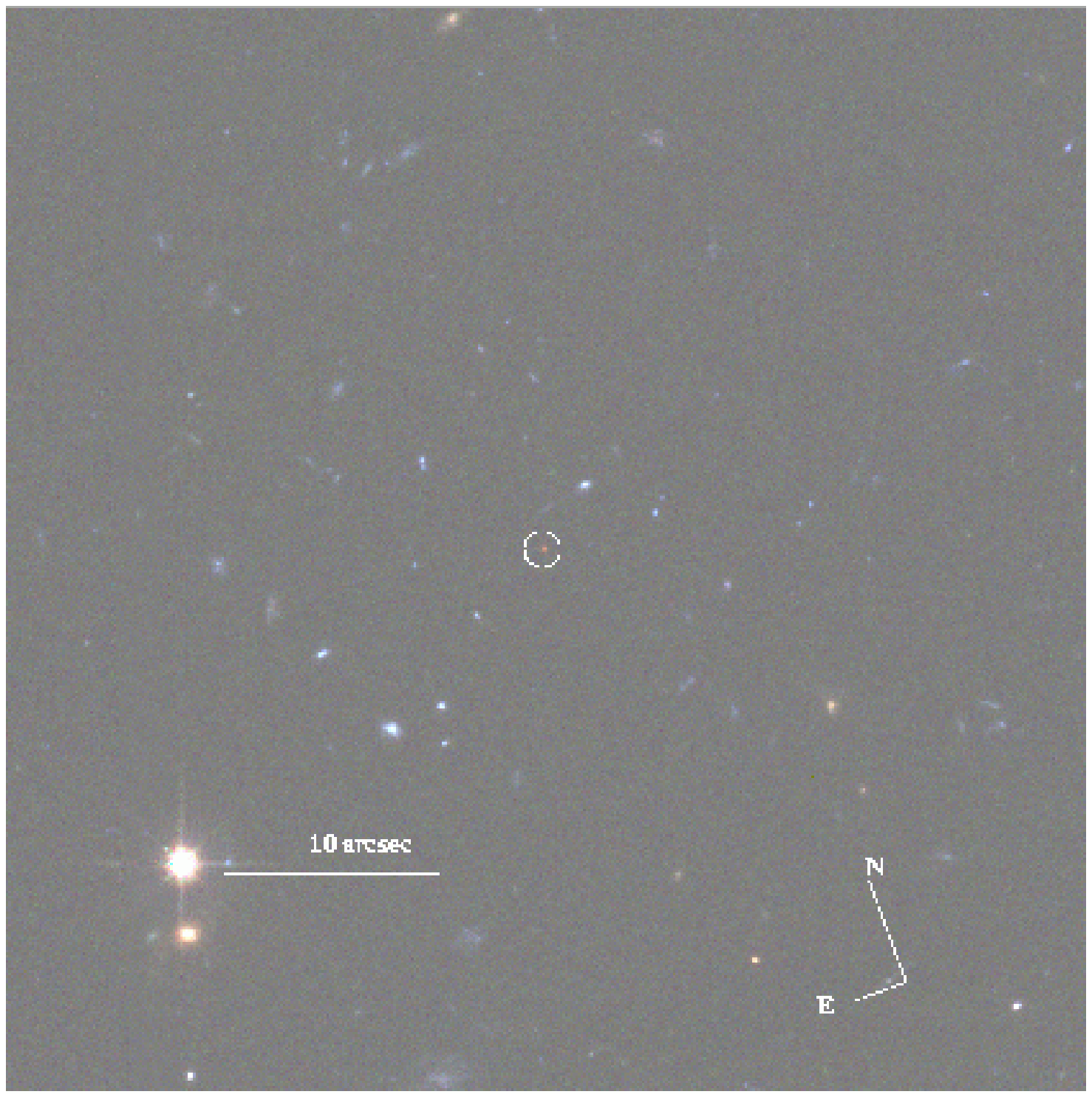}}
\end{minipage}
\begin{minipage}{0.165\textwidth}
\resizebox{\textwidth}{!}{\includegraphics{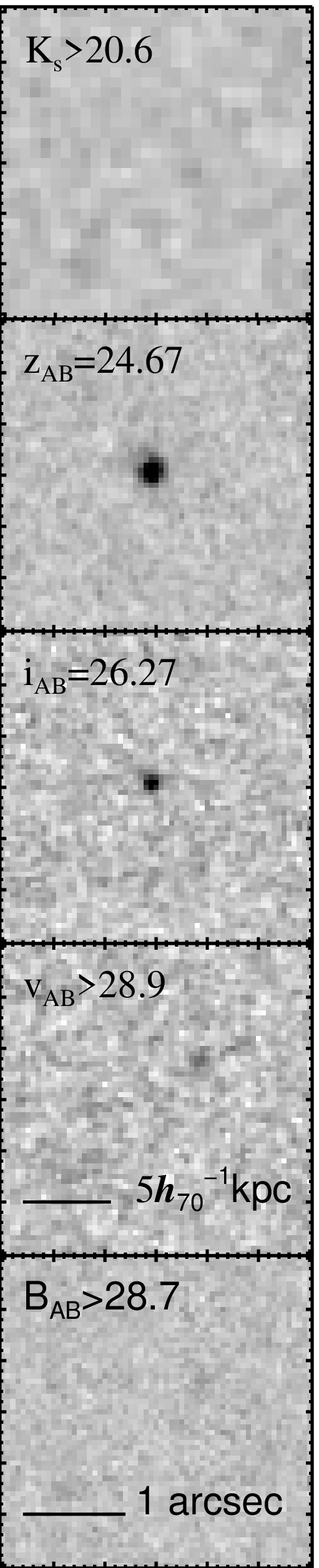}}
\end{minipage}
\caption{Three-colour picture from {\em HST}/ACS GOODS imaging of the
Chandra Deep Field South, formed from the first 3 epochs of
observation. Red channel is the F850LP $z'$-band (6\,ksec), green is
the F775W $i'$-band (3\,ksec) and blue is the F606W $v$-band
(3\,ksec). The $z=5.78$ galaxy SBM03\#3 is at the centre, and is the
reddest object in the field (the only $i'$-drop). The field of view is
$50''\times 50''$. The $B$, $v$, $i'$, $z'$
\& $K_{\rm s}$ images are shown on the right (3\,arcsec across).}
\label{fig:colpic}
\end{figure*}

\end{document}